\documentclass[twocolumn]{aastex61}

\usepackage{color}
\received{Jan 30, 2018}
\revised{Mar 02, 2018}
\accepted{Mar 04 2018}
\submitjournal{ApJL}

\shorttitle{A hydrated asteroid in the Kuiper Belt.}
\shortauthors{Seccull et al.}

\begin{document}

\title{2004 EW$_{95}$: A phyllosilicate bearing carbonaceous asteroid in the Kuiper Belt.}

\correspondingauthor{Tom Seccull}
\email{tseccull01@qub.ac.uk}

\author[0000-0001-5605-1702]{Tom Seccull}
\affil{Astrophysics Research Centre, Queen's University Belfast, Belfast, BT7 1NN, UK}

\author[0000-0001-6680-6558]{Wesley C. Fraser}
\affiliation{Astrophysics Research Centre, Queen's University Belfast, Belfast, BT7 1NN, UK}

\author[0000-0003-0350-7061]{Thomas H. Puzia}
\affiliation{Institute of Astrophysics, Pontificia Universidad Cat\'olica de Chile, Av. Vincu\~na Mackenna 4860, 7820436, Santiago, Chile}

\author[0000-0002-8255-0545]{Michael E. Brown}
\affiliation{Division of Geological and Planetary Sciences, California Institute of Technology, Pasadena, CA 91125, USA}

\author{Frederik Sch\"onebeck}
\affiliation{Astronomisches Rechen-Institut, Zentrum f\"ur Astronomie der Universit\"at Heidelberg, M\"onchhofstra{\ss}e 12-14, 69120 Heidelberg, Germany}



\begin{abstract}

Models of the Solar System's dynamical evolution predict the dispersal of primitive 
planetesimals from their formative regions amongst the gas-giant planets due to the 
early phases of planetary migration. Consequently, carbonaceous objects were 
scattered both into the outer asteroid belt and out to the Kuiper Belt. These 
models predict that the Kuiper Belt should contain a small fraction of objects with 
carbonaceous surfaces, though to date, all reported visible reflectance spectra of 
small Kuiper Belt Objects (KBOs) are linear and featureless. We report the unusual
reflectance spectrum of a small KBO, (120216) 2004~EW$_{95}$, exhibiting a 
large drop in its near-UV reflectance and a broad shallow optical absorption
feature centered at $\sim700$~nm which is detected at greater than 
4$\sigma$ significance. These features, confirmed through multiple epochs of 
spectral photometry and spectroscopy, have respectively been associated with
ferric oxides and phyllosilicates. The spectrum bears striking resemblance 
to those of some C-type asteroids, suggesting that 2004~EW$_{95}$ may share a
common origin with those objects. 2004~EW$_{95}$ orbits the Sun in a stable mean 
motion resonance with Neptune, at relatively high eccentricity and inclination, 
suggesting it may have been emplaced there by some past dynamical instability. 
These results appear consistent with the aforementioned model predictions and are
the first to show a reliably confirmed detection of silicate material on a small KBO.

\end{abstract}




\section{Introduction} \label{sec:intro}

Current models of the Solar System's dynamical evolution, such as the Nice model
\citep{2005Natur.435..466G,2005Natur.435..462M,2005Natur.435..459T,2011AJ....142..152L}, 
predict that the Kuiper Belt should largely be composed of objects which formed 
beyond the giant planet region.  Concurrently, the Grand Tack model 
\citep{2011Natur.475..206W,2012M&PS...47.1941W} posits that the primitive 
carbonaceous asteroids formed amongst the giant planets and were injected 
into the outer asteroid belt due to the early migrations of Jupiter and Saturn. By 
the same mechanism the Grand Tack model also predicts that carbonaceous asteroids 
could have been scattered outward into the Kuiper Belt region, suggesting that a 
small number of objects beyond Neptune would possess primitive, dark, 
carbon rich, asteroidal surfaces like those of C/D/P-type asteroids.

The optical (400 $ \leq \lambda \leq$ 900 nm) reflectance spectra of small KBOs 
typically exhibit a red, linear, featureless slope that reveals little about their 
surface composition \citep{2009A&A...508..457F}. So far the only material commonly 
found on the surfaces of small KBOs is water ice, which characteristically absorbs 
in the NIR at $\sim 1.5 \mbox{ $\mu$m}$ and $\sim2.0 \mbox{ $\mu$m}$ 
\citep{2011Icar..214..297B,2012AJ....143..146B}. No diagnostic optical features 
have been confirmed.

(120216) 2004~EW$_{95}$ is a 
small 3:2 resonant KBO with absolute magnitude, $H_{R}~=~6.31~\pm~0.05$ 
\citep{2015A&A...577A..35P} and radiometric diameter, $r~=~291.1_{-25.9}^{+20.3}~ 
km$ \citep{2012A&A...541A..93M}. Its reflectance spectrum was initially revealed to 
be atypical of KBO spectra from optical and NIR spectral photometry gathered with 
the Hubble Space Telescope (HST) \citep{2015ApJ...804...31F}. Of the eight small 
outer Solar System objects observed in HST GO-Program 12234, 
2004~EW$_{95}$ was unique in that instead of exhibiting a linear optical spectrum, 
it possessed an apparent upward curvature through the optical range. This peculiar 
property of the object, along with hints of a drop in its near-UV reflectance 
\citep{2015ApJ...804...31F}, warranted subsequent spectroscopic observations.

\section{Observations} \label{sec:obs}



We observed 2004 EW$_{95}$ twice at the European Southern Observatory's Very Large
Telescope, during 2014 August 3 and 2017 April 22-23 using the X-Shooter
\citep{2011A&A...536A.105V} and FORS2 \citep{1998Msngr..94....1A} spectrographs
respectively.

X-Shooter is a medium resolution echelle spectrograph with three arms covering the 
near-UV/blue (UVB; 300---560 nm), visual (VIS; 550---1020 nm) and NIR (1024---2480 
nm) spectral range. While 2004~EW$_{95}$ was observed in all three of X-Shooter's 
arms simultaneously, the NIR observations had such a low SNR that they could not be 
properly reduced. Hence from this point we only consider the UVB and VIS 
observations. The UVB and VIS detectors share a common pixel scale of 
0.16\arcsec~and slit length of 11\arcsec. We set the UVB and VIS slits to widths of 
1.0\arcsec~and 0.9\arcsec, each providing a respective resolving power of 
$\sim$5100 and $\sim$8800. All observations were performed with a UVB and VIS 
detector readout binning of 1x2, except for observations of the solar calibrator HD 
117286, when no binning was performed. The difference in binning between 
2004~EW$_{95}$ and HD 117286 produces no observable change to the features we 
detect in the UVB and VIS reflectance spectrum when compared to those calibrated 
with the other stars. 

The X-Shooter 
observations were performed in a 3 point dither pattern to mitigate 
bad pixel artifacts and contamination by cosmic rays in the stacked spectrum. The 
slit was realigned to the parallactic angle at the beginning of each observation to 
reduce the effects of atmospheric differential refraction; this was especially 
important in light of X-Shooter's disabled Atmospheric Dispersion Corrector (ADC) 
at the time of observing. Of 9 total observed exposures of 2004~EW$_{95}$, 7 were 
usable. Two exposures where the object appeared to drift off the slit were 
ignored during data reduction. Three solar calibrator stars, HD~117286, Hip~075235 
and Hip~077439 were similarly observed adjacent in time to 2004~EW$_{95}$ and at 
similar airmass. Flux calibrators EG~274 and Feige~110 were observed as part of the 
standard calibration program. Details of observing conditions are reported in Table 
\ref{tab:obsdets}.

\begin{table*}
\begin{center}
\caption{Spectroscopy Observation Details}
\label{tab:obsdets}
\begin{tabular}{ p{2.4cm}p{5cm}p{0.8cm}p{0.8cm}p{0.8cm}p{2cm}p{2cm}}
\hline\hline
Target & Observation Date $\vert$ UT Time & \multicolumn{3}{c}{Exposure Time (s)} & Airmass & Seeing (\arcsec) \\[2pt]
\hline
FORS2 Night 1 & & & & & &\\
\hline
\textbf{BD-00 2514} & 2017-04-22 $\vert$ 01:30:52 - 01:41:46 & 12.0 & & & 1.126 - 1.127 & 0.51 - 0.51 \\[1pt]
TYC-4949-897-1 & 2017-04-22 $\vert$ 02:43:21 - 02:50:52 & 12.0 & & & 1.106 - 1.107 & 0.57 - 0.59 \\[1pt]
HD 106436 & 2017-04-22 $\vert$ 02:57:45 - 03:07:26 & 6.0 & & & 1.005 - 1.005 & 0.67 - 0.69 \\[1pt]
HD 106649 & 2017-04-22 $\vert$ 03:07:30 - 03:13:23 & 3.0 & & & 1.003 - 1.003 & 0.56 - 0.56 \\[1pt]
2004 EW95 & 2017-04-22 $\vert$ 03:26:06 - 04:10:18 & 500.0 & & & 1.039 - 1.092 & 0.54 - 0.66 \\[1pt]
HD 136122 & 2017-04-22 $\vert$ 04:10:54 - 04:19:18 & 3.0 & & & 1.087 - 1.087 & 0.54 - 0.59 \\[1pt]
HD 140854 & 2017-04-22 $\vert$ 04:19:24 - 04:27:41 & 3.0 & & & 1.125 - 1.125 & 0.58 - 0.58 \\[1pt]
\hline
FORS2 Night 2 & & & & & &\\
\hline
HD 106436 & 2017-04-23 $\vert$ 02:13:57 - 02:21:40 & 6.0 & & & 1.012 - 1.012 & 0.65 - 0.68 \\[1pt]
HD 106649 & 2017-04-23 $\vert$ 02:22:15 - 02:30:42 & 3.0 & & & 1.005 - 1.005 & 0.57 - 0.57 \\[1pt]
2004 EW95 & 2017-04-23 $\vert$ 02:37:58 - 03:27:35 & 500.0 & & & 1.116 - 1.183 & 0.60 - 0.72 \\[1pt]
HD 136122 & 2017-04-23 $\vert$ 03:36:06 - 03:46:12 & 3.0 & & & 1.149 - 1.149 & 0.71 - 0.72 \\[1pt]
HD 140854 & 2017-04-23 $\vert$ 03:46:22 - 03:49:38 & 3.0 & & & 1.203 - 1.206 & 0.62 - 0.69 \\[1pt]
\hline
\hline
X-Shooter & & UVB & VIS & NIR & &\\
\hline
\textbf{HD 117286} & 2014-08-02 $\vert$ 23:42:36 - 23:48:40 & 3.0 & 3.0 & 3.0 & 1.170 - 1.188 & 1.25 - 1.60 \\[1pt]
2004 EW95 & 2014-08-03 $\vert$ 00:21:15 - 01:55:31 & 500.0 & 466.0 & 532.0 & 1.166 - 1.715 & 0.99 - 1.30 \\[1pt]
HIP 077439 & 2014-08-03 $\vert$ 02:08:38 - 02:12:04 & 3.0 & 3.0 & 3.0 & 1.331 - 1.347 & 1.33 - 1.67 \\[1pt]
HIP 075235 & 2014-08-03 $\vert$ 02:19:27 - 02:23:05 & 3.0 & 3.0 & 3.0 & 1.568 - 1.579 & 1.19 - 1.69 \\[1pt]
\hline
\end{tabular}\\[6pt]
\small{\textsc{Note---} Stars used to calibrate the spectra in Figs. 2 \& 3 are highlighted in bold.}
\end{center}
\end{table*}

FORS2 is a multipurpose spectrograph and imager. 
We obtained FORS2 spectra of 2004~EW$_{95}$ over two nights in Long Slit 
Spectroscopy mode with a 2 point dither pattern, using the blue sensitive E2V 
detector, the standard resolution collimator, and the 600B+22 grism. At the 
beginning of each image pair the slit was realigned to the parallactic angle. On 
each night 4 exposures were obtained for 2004~EW$_{95}$. In addition, on each night 
two solar calibrators were observed just before and just after 2004~EW$_{95}$. 
These calibrators were: HD106436, HD 106649, HD 136122, and HD140854. Two 
additional solar calibrators (BD-00 2514 \& TYC-49494-897-1) were observed two 
hours prior to 2004 EW$_{95}$ on night 1. All the solar calibrators were observed 
at similar airmass to 2004~EW$_{95}$. BD-00 2514, the star that produced the lowest 
calibration residuals below 400 nm and at ~517 nm was used to calibrate the science 
spectra presented in \S\ref{sec:res}. Flux standards were observed but the images 
were later found to be saturated and therefore useless. During the reduction 
process all of the FORS2 spectra were corrected for airmass extinction using the 
instrument extinction table. Over both nights the cloud cover never warranted 
greater than a "Clear" designation and does not appear to have affected the slope 
of the final spectra. Further details on the FORS2 observations are reported in 
Table \ref{tab:obsdets}.

\section{Data Reduction} \label{sec:red}

Standard reduction steps (including order rectification and merging, and flux 
calibration in the case of X-Shooter) were performed for all the observed spectra 
with the European Southern Observatory (ESO) provided instrument pipelines 
\citep[X-Shooter v.2.7.1; FORS2 v.5.3.2,][]{2017MAN..XSHOOPIPE..M,2017MAN..FORS2PIPE..S} 
in the ESO Reflex data processing environment (v. 2.8.4 and v. 2.8.5 respectively) 
\citep{2013A&A...559A..96F}.

We applied two different methods for sky subtraction, cosmic ray removal and 
extraction of the spectra. This was done to test the consistency of the two methods 
and confirm that any observed features were not simply reduction artifacts.

\subsection{Method 1} \label{sec:meth1}

\begin{figure}
\plotone{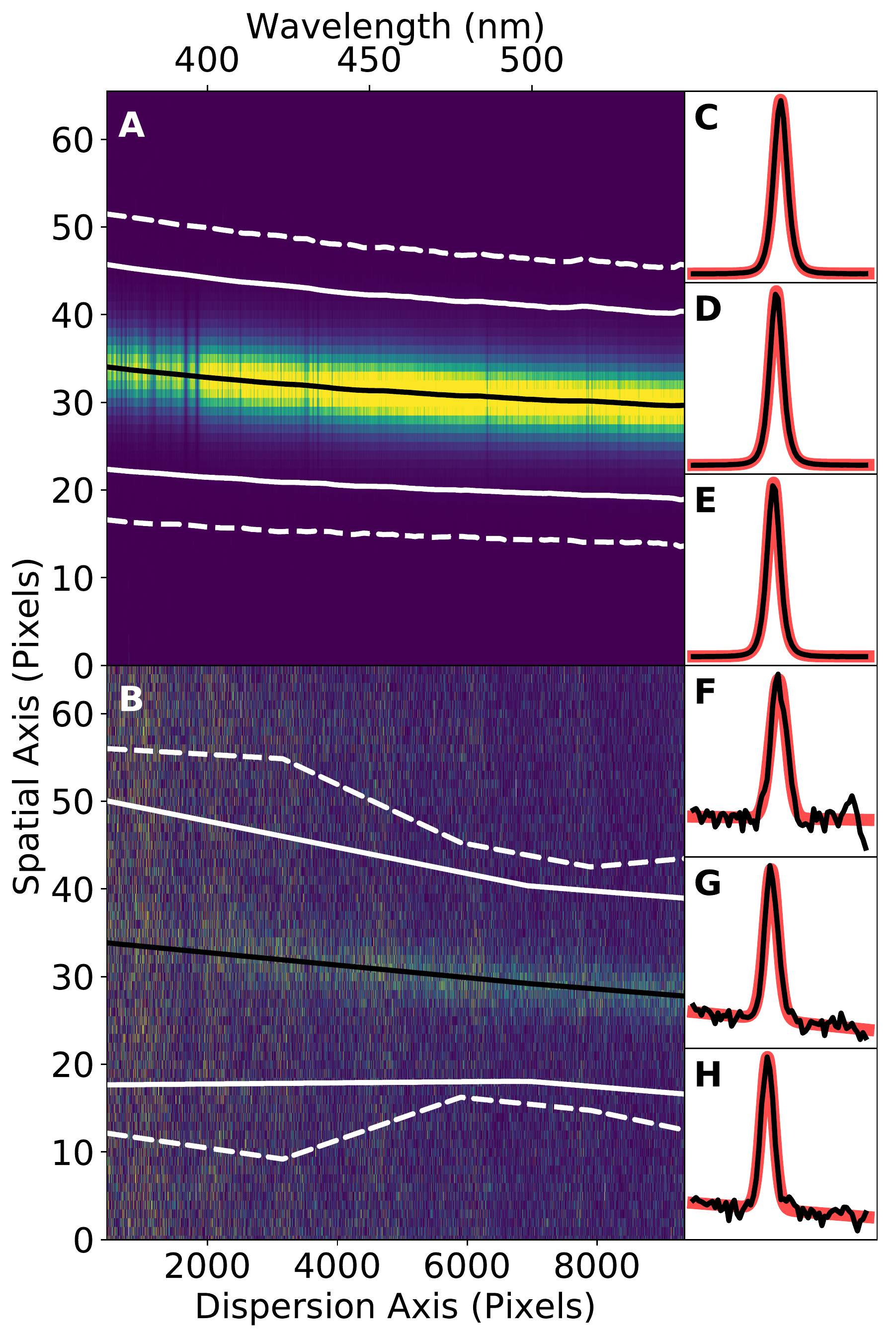}
\caption{Diagram of the method 1 reduction process for the 2D flux calibrated, 
rectified and merged UVB X-Shooter spectra of solar calibrator star HD 117286 
(panels A, C--E) and 2004~EW$_{95}$ (panels B, F--H). The amount of flux gathered 
for HD 117286 in panel A is $\sim$5 orders of magnitude greater than that gathered 
for 2004~EW$_{95}$ in panel B. Panels A and B show lines tracing the 
sky subtraction limits (dashed white) used to subtract the sky flux from these 
images. Also shown are lines tracing the Moffat profile centers for each extraction 
bin along the dispersion axis (solid black), and their associated extraction limits 
(solid white). Panels C--H show the median spatial profiles of example data bins 
taken from various wavelengths along the spectrum (black), fitted with their 
associated Moffat profiles (red).  \label{fig:f1}}
\end{figure}

Due to X-Shooter's disabled ADC at the time we observed 2004 EW$_{95}$, 
the spectra were found to have a wavelength dependent spatial position in the 2D 
rectified images, which was most pronounced at the shortest wavelengths (see Fig. 
\ref{fig:f1}). The Point Spread Function (PSF) of the spectrum was also wavelength 
dependent, due to a combination of variations across X-Shooter's echelle orders and 
wavelength dependent seeing. These factors made a simple straight extraction of the 
1D spectrum impossible without including increased background noise in the 
extracted spectrum. A Python script was created to track the wavelength dependent 
spectrum width and center within each image to evaluate wavelength-dependent
extraction limits, thus preserving Signal to Noise Ratio 
(SNR) while avoiding wavelength-dependent extraction losses, especially in the 
near-UV. The script worked as follows.

To first remove the highly variable background level, each 2D pipeline reduced and 
flux calibrated spectrum was binned along the dispersion axis. The bins were 
progressively widened until the SNR of the summed spatial profile reached a 
predetermined threshold. Moffat profiles \citep{1969A&A.....3..455M} were fitted to 
the median spatial profile of each bin in order to define the science extraction 
limits, and sky regions. Sky region boundaries were taken as the pixels outside 
$\pm3$ Full-Widths at Half Maximum (FWHM) from each Moffat 
profile center. These boundaries were then linearly interpolated across the full 
unbinned image. In each unbinned wavelength element the background value was taken 
as the median value in the sky regions and was subtracted from the image at that 
wavelength. Cosmic rays in the sky regions and target regions were then 
separately sigma clipped at 5$\sigma$. The binning and fitting process was repeated 
on the background subtracted images to define the target extraction limits on the 
sky subtracted image. The science extraction 
limits were set to $\pm2$ FWHM from each Moffat profile center (see 
Fig. \ref{fig:f1}).  The flux within the interpolated extraction limits was summed 
for each unbinned wavelength element to extract the 1D spectrum. Dithers in which 
the FWHM extraction boundaries fell off the image were omitted from later stacking.

\subsection{Method 2} \label{sec:meth2}

The spectra were reduced using only the ESO Reflex instrument pipelines 
\citep{2013A&A...559A..96F} which performed the sky subtraction, cosmic ray 
flagging and spectrum extraction as described in the pipeline user manuals 
\citep{2017MAN..XSHOOPIPE..M,2017MAN..FORS2PIPE..S}. The FORS2 spectra were 
extracted with both optimal \citep{1986PASP...98..609H} and aperture methods, while 
the X-Shooter spectra were extracted with only the aperture method. The widths of 
the straight extraction apertures for each spectrum were set at the greatest 
separation between the extraction limits calculated using method 1.

\begin{figure*}
\plotone{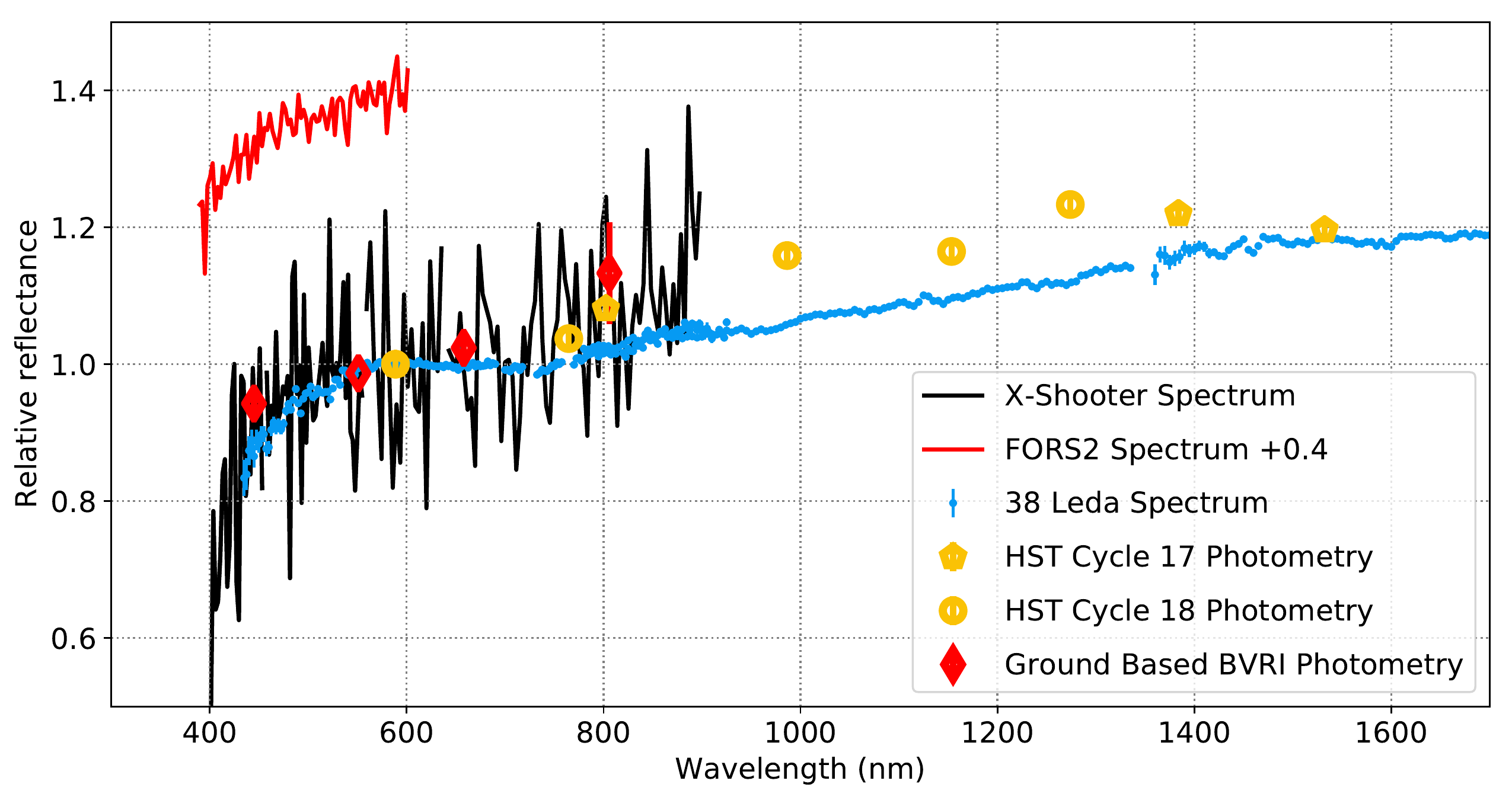}
\caption{Reflectance spectra and photometry \citep{2015ApJ...804...31F} of 
2004~EW$_{95}$ compared to the combined optical and NIR reflectance spectrum of the 
hydrated C-type asteroid, 38 Leda from the SMASSII and SMASSir catalogs 
\citep{2002Icar..158..106B,2002Icar..158..146B,2009Icar..202..160D}. 
2004~EW$_{95}$'s drop in reflectance toward the UV is clearly visible in both the 
X-Shooter and FORS2 spectra, matching well with 38 Leda. The 
presence of the broad feature centered near 700 nm is apparent in both 
the X-Shooter spectrum and the HST spectrophotometry. We attribute this feature to 
phyllosilicate absorption like that of the hydrated C-type asteroids. The NIR 
behavior observed for 2004~EW$_{95}$ in the HST photometry closely resembles 
the NIR spectral behavior of C-type asteroids, presenting a featureless red slope, 
remaining roughly constant from $\sim$1000 nm to $\sim$1400 nm. 2004 EW$_{95}$'s 
reflectance drops slightly at $\sim$1500 nm hinting at possible absorption due to 
surface water ice. Reflectances in all datasets are normalized at 589 nm. The FORS2 
spectrum (in red) is offset by +0.4 for clarity. The apparent difference in 
overall slope between overlapping regions of the FORS2 and X-Shooter spectra 
are calibration artifacts resulting from the use of slightly different solar 
analogue stars. \label{fig:f2}}
\end{figure*}

All methods tested produced spectra with consistent features for both FORS2 and 
X-Shooter observations. However, spectra extracted via method 1 showed increased 
SNR relative to those reduced via method 2. For this reason the spectra presented 
in \S\ref{sec:res} were extracted via method 1.

Following extraction, the individual spectra were normalized, median stacked, solar 
calibrated, and binned. Dithers with extremely low SNR were omitted from the final 
stack. Regions near $\sim$456, $\sim$560 and $\sim$640 nm in the X-Shooter spectrum 
were rejected to avoid copious artifacts produced by bad pixels and the edges of 
the echelle orders.

Since the UVB arm and VIS arm of the X-Shooter spectrum were normalized at 
different wavelengths during stacking, the arms required rescaling relative to each 
other to produce a continuous spectrum. The scaling factor of UVB relative to VIS 
was calculated as the ratio between the relative flux of the KBO with respect to 
the flux of the solar calibrator star measured at the wavelengths of normalization 
in each arm. The scaling was then adjusted to account for the spectral slope in the 
region where the UVB and VIS arms join.

In all spectra the shortest wavelength has been limited to $\sim$400 nm due to 
large residuals produced by differences in metallicity and temperature between the 
solar calibrators used and the Sun \citep{1980A&A....91..221H}.

The FORS2 spectrum presented in Figures \ref{fig:f2} \& \ref{fig:f3} is comprised 
of observations only from night 1, observed under photometric conditions. 

To further display the integrity of our reduction methods we show the spectrum of 
the KBO 1999~OX$_3$ in Figure~\ref{fig:f3}. The X-Shooter data of this target 
exhibit a very similar Signal to Noise as in the X-Shooter data of 2004~EW$_{95}$. 
Thus, any extraction issues of our pipeline that may be apparent in the 
spectra of 2004~EW$_{95}$, should be equally apparent in the spectra of 
1999~OX$_3$. As can be seen in Figure~\ref{fig:f3}, the spectra produced 
with extraction method 1 result in a typical KBO spectrum, that is linear, 
featureless, red and exhibits no identifiable absorption features. Hence we 
conclude that the features observed in the spectrum of 2004~EW$_{95}$ are not 
reduction artifacts and instead are inherent to the spectrum itself. Via linear 
regression, the optical slope of 1999~OX$_3$'s spectrum was measured to be 30.6 
$\pm$ 1.5\% per 100 nm, in accord with literature values 
\citep{2015A&A...577A..35P}. 

\section{Results \& Discussion} \label{sec:res}

We have detected two features in the reflectance spectrum of 3:2 resonant KBO 
2004~EW$_{95}$: a large drop in reflectance at wavelengths below 550 nm, and a 
broad ($\sim$0.2--0.3 $\mu$m wide), shallow absorption feature centered at around 700 nm 
(see Fig. \ref{fig:f2}). Previously, neither feature has ever been reliably 
detected in the spectrum of a KBO. Each of these two features has been 
independently observed by two separate instruments, and are present in the reported 
reflectance spectra regardless of which of our solar calibrator targets, or which 
of two separate spectral extraction techniques were used in our data reduction. 

\begin{figure}
\plotone{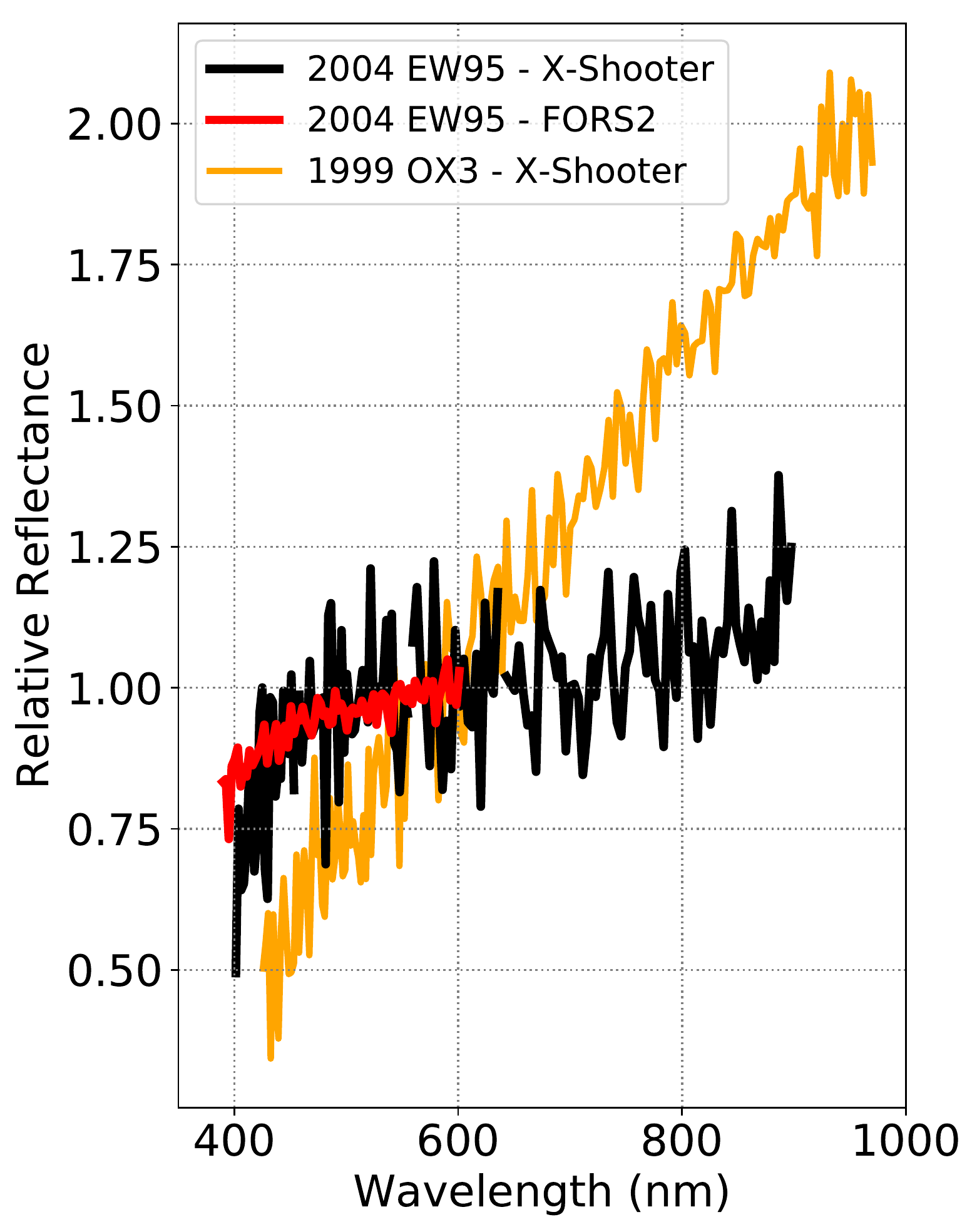}
\caption{Spectrum comparison plot. Here we compare the spectra of 2004~EW$_{95}$ observed with X-Shooter and FORS2, along with the X-Shooter spectrum  of a typical KBO, 1999~OX$_3$. Both spectra of 2004~EW$_{95}$ agree with each other very well at 
wavelengths greater than $\sim$430 nm. Below $\sim$430 nm there is a divergence 
in the slope of each spectrum caused by difference in color of the solar 
calibrators used. Both 2004~EW$_{95}$ and 1999~OX$_3$ were of similar brightness 
when observed. All spectra have been reduced using the method 1 
extraction technique described in \S\ref{sec:red}. \label{fig:f3}}
\end{figure}

The FORS2 and X-Shooter spectra of 2004~EW$_{95}$  fully agree 
above 430 nm (see Fig. \ref{fig:f3}). At 415 nm the X-Shooter spectrum has a 
greater spectral slope and is discrepant from the FORS2 spectrum in that wavelength 
range by $\sim$2$\sigma$. This difference corresponds to the known difference in 
\bv ~color between the solar calibrator stars used to calibrate each spectrum.

The UV-optical spectrum of 2004~EW$_{95}$ bears a striking resemblance to those of 
primitive carbonaceous asteroids. Specifically 2004~EW$_{95}$ resembles a hydrated 
C-type asteroid \citep{2002Icar..158..146B,2009Icar..202..160D,2016AJ....152...54V} 
(we compare 2004~EW$_{95}$ to the asteroid 38 Leda in Figure \ref{fig:f2}). The 
shallow optical absorption feature at $\sim$700~nm defines the hydrated Ch/Cgh 
asteroid subclasses, being observed in $\sim$30\% of C-types 
\citep{2012Icar..221..744R}, and has been associated with 
charge transfer in aqueously altered silicate material 
\citep{1989Sci...246..790V,2014Icar..233..163F,2015aste.book...65R}. To 
characterize the $\sim700$~nm feature and test the significance of its detection, 
we first remove a linear continuum slope of 3.6\% per 100~nm that was determined 
from a linear fit to the 530--580~nm and 850--900~nm wavelength ranges. A Gaussian 
profile was then fit to the continuum in a maximum likelihood sense, using the 
emcee Monte-Carlo Markov Chain sampler \citep{2013PASP..125..306F}. We adopt as a 
best-fit the median sample point, uncertainty as the $1\sigma$ sampling range when 
marginalizing over the other parameters. The best-fit depth, center, and FWHM were 
$4^{+1}_{-1}\%$, $734^{+43}_{-45}$~nm and $319^{+101}_{-101}$~nm, respectively. The 
$4\sigma$ lower limit on the feature depth is 1\% demonstrating the veracity of 
the detection. Technically, our routine quoted a higher significance of detection, 
however at this high a significance, the ability to determine the continuum is the 
ultimate limiting factor.

Indications for phyllosilicate features have previously been reported in the 
spectra of KBOs 2003~AZ$_{84}$ \citep{2004A&A...421..353F}, 2000~EB$_{173}$ (38628 
Huya) and 2000~GN$_{171}$ \citep{2003AJ....125.1554L,2004A&A...416..791D}. Follow-up spectra of sufficient quality to detect those reported features, however, have
revealed featureless spectra on later occasions, with rotational spectral 
variability reported as a possible 
but unconfirmed explanation for the disappearance of the feature on these objects 
\citep{2004A&A...421..353F,2009A&A...508..457F,2017A&A...604A..86M}. Repeat 
photometric observations of 2004~EW$_{95}$ with HST reported by 
\citet{2015ApJ...804...31F} have shown its spectrum to be invariable, which is 
supported by the consistency between that HST spectrophotometry of 2004~EW$_{95}$ 
and the reflectance spectra reported in this work (see Fig. \ref{fig:f2}). With 
multiple independent photometric and spectroscopic detections of the 700 nm feature 
in the spectrum of 2004~EW$_{95}$ over multiple epochs, we report the first confident 
detection of phyllosilicates on any KBO.

Like 2004~EW$_{95}$, some S, V and C-type asteroids also exhibit a drop in near-UV 
reflectance, including the hydrated C-types to which this spectrum is most similar. 
On asteroids the UV drop has been attributed to the presence of ferric oxide 
material \citep{2002Icar..158..106B}. It should be noted that other materials such 
as complex aromatic organics can also produce a similar drop-off 
\citep{2014Icar..237..159I,2016M&PS...51..105H}, though organic rich bodies such as 
the P \& D-type asteroids, have not been observed to exhibit this UV drop in their 
reflectance spectra \citep{2002Icar..158..106B,2002Icar..158..146B,2009Icar..202..160D,2014A&A...568L...7M}. 
Moreover, other than 2004~EW$_{95}$ of the 41 published optical spectra of KBOs and 
centaurs with sufficient short wavelength coverage, only the centaur (32532) 
Thereus hints at the presence of a similar UV drop \citep{2002A&A...392..335B} 
though the presence of this feature on Thereus has not yet been confirmed.

The longest wavelength at which 2004~EW$_{95}$ was observed, was in the HST/Wide 
Field Camera 3 153M filter centered at 1532.2 nm (see Figure \ref{fig:f2}). Here the observed reflectance of the object appears to decrease relative to the 
NIR photometric points from 1000---1400~nm. The decrease is consistent with the 
presence of a small amount of water ice, which characteristically absorbs at these 
wavelengths \citep{2012AJ....143..146B,2012ApJ...749...33F}. This feature is the 
only one to distinguish the spectrum of 2004~EW$_{95}$ from that of a C-type 
asteroid and suggests that unlike the C-types in the outer asteroid belt, 
2004~EW$_{95}$ has retained its primordial surface water content.

The Grand Tack model \citep{2011Natur.475..206W,2012M&PS...47.1941W} predicts that 
as a result of the early migrations of Jupiter and Saturn, the primitive small 
bodies that formed between the gas giants would be scattered, and injected into the 
outer asteroid belt, making up the bulk of the organic-rich asteroids. 
Models of Jupiter's and Saturn's rapid gas accretion show that 
primitive interplanetary asteroids could also be scattered as the gas giants formed 
\citep{2017Icar..297..134R}. By either mechanism (they are not mutually 
exclusive), a small fraction of those bodies would be scattered outward into the 
Trans-Neptunian belt, where they could later be captured into the Mean Motion 
Resonances (MMRs) of Neptune \citep{2008Icar..196..258L}. 2004~EW$_{95}$ orbits the 
Sun in Neptune's 3:2 MMR, at relatively high orbital eccentricity and inclination 
\citep[a = 39.316 AU, e = 0.3139, i = 29.3$\degr$,]
[]{2015A&A...577A..35P}. The presence of a phyllosilicate feature 
indicates that 2004~EW$_{95}$ has been subjected to significant heating, either 
radiogenic \citep{2015Icar..245..320M}, from a very large single collision or 
extensive collisional bombardment \citep{1995Icar..113..156R, 2002ESASP.500...29M}, 
or via solar irradiation. The striking similarity between 2004~EW$_{95}$ and 
certain C-type asteroids points to the plausible idea that 2004~EW$_{95}$ shares a 
common origin with these objects. Taken together, the spectroscopic similarity to 
C-type asteroids and the orbital properties of 2004~EW$_{95}$ are consistent with 
the idea that this object may have formed near Jupiter amongst the primordial 
C-type asteroids \citep{2011Natur.475..206W} and was subsequently emplaced into the 
Kuiper Belt by the migrating planets.

\acknowledgments
We thank Faith Vilas and Alan Fitzsimmons for their encouraging 
constructive discussion and comments. This work is based on observations collected 
at the European Organisation for Astronomical Research in the Southern Hemisphere 
under ESO programmes 093.C-0259(A), 095.C-0521(A) and 099.C-0651(A). W.C.F. 
acknowledges support from STFC grant ST/P0003094/1. T.H.P. acknowledges support 
through the FONDECYT Regular Project No. 1161817 and the BASAL Center for 
Astrophysics and Associated Technologies (PFB-06). M.E.B. acknowledges support from 
the NASA Planetary Astronomy Program through Grant NNX09AB49G.

%

\vspace{5mm}
\facilities{ESO VLT(X-Shooter and FORS2)}


\software{\\
Astropy \citep{2013A&A...558A..33A}\\
corner \citep{2016JOSS....1...24F}\\
emcee \citep{2013PASP..125..306F}\\
ESO Reflex \citep{2013A&A...559A..96F}\\
matplotlib \citep{2007CSE...9..3H}\\
numpy \citep{2011CSE...13..22V}\\
scipy \citep{2001SCIPY..J}
}





\end{document}